\documentclass[journal,twoside,web]{ieeecolor}
\usepackage{etoolbox}
\makeatletter
\@ifundefined{color@begingroup}%
{\let\color@begingroup\relax
\let\color@endgroup\relax}{}%
\def\fix@ieeecolor@hbox#1{%
\hbox{\color@begingroup#1\color@endgroup}}
\patchcmd\@makecaption{\hbox}{\fix@ieeecolor@hbox}{}{\FAILED}
\patchcmd\@makecaption{\hbox}{\fix@ieeecolor@hbox}{}{\FAILED}

\let\labelindent\relax
\usepackage{tmi}
\usepackage{cite}
\usepackage{amsmath,amssymb,amsfonts}

\usepackage{algorithmic}
\usepackage{dsfont}
\usepackage{textcomp}
\usepackage{multirow}
\usepackage{enumitem}
\usepackage{amsmath}
\usepackage{pifont}
\usepackage{fontawesome}
\usepackage[pdftex]{graphicx}
\usepackage{threeparttable}

\usepackage{amsmath}
\usepackage{amssymb}
\usepackage{algorithmic}
\usepackage{csquotes}
\usepackage{color, colortbl}
\definecolor{Gray}{gray}{0.9}
\definecolor{Gray2}{gray}{0.95}
\usepackage{amsmath}
 \usepackage[ruled]{algorithm2e}    
 
\newcommand{\etal}{\textit{et al.}}
\begin{document}

\title{Prompt-driven Latent Domain Generalization for Medical Image Classification}
\author{Siyuan Yan, Chi Liu, Zhen Yu, Lie Ju, Dwarikanath Mahapatra, Brigid Betz-Stablein, Victoria Mar, Monika Janda, Peter Soyer, and Zongyuan Ge, \IEEEmembership{Senior Member, IEEE}
\thanks{S. Yan, C. Liu, Z. Yu, L. Ju and Z. Ge is with the Monash University, Clayton, VIC. 3800 Australia, and also with Airdoc-Monash Research, Monash University, Clayton, VIC. 3800 Australia (E-mail: siyuan.yan@monash.edu, zongyuan.ge@monash.edu).}
\thanks{D. Mahapatra is with the Inception Institute of AI, Abu Dhabi, UAE (E-mail: dwarikanath.mahapatra@inceptioniai.org)}
\thanks{B. Betz-Stablein, M. Janda and P. Soyer is with the University of Queensland Diamantina Institute, Dermatology Research Centre,
The University of Queensland, Brisbane, Australia (E-mail: ,p.soyer@uq.edu.au).}
\thanks{V. Mar is with Victorian Melanoma Service, Alfred Health, Melbourne, VIC. 3004, Australia (Email: victoria.mar@monash.edu )}
}

\maketitle

\begin{abstract}
Deep learning models for medical image analysis easily suffer from distribution shifts caused by dataset artifacts bias, camera variations, differences in the imaging station, etc., leading to unreliable diagnoses in real-world clinical settings. Domain generalization (DG) methods, which aim to train models on multiple domains to perform well on unseen domains, offer a promising direction to solve the problem. However, existing DG methods assume domain labels of each image are available and accurate, which is typically feasible for only a limited number of medical datasets. To address these challenges, we propose a novel DG framework for medical image classification without relying on domain labels, called Prompt-driven Latent Domain Generalization (PLDG). PLDG consists of unsupervised domain discovery and prompt learning. This framework first discovers pseudo domain labels by clustering the bias-associated style features, then leverages collaborative domain prompts to guide a Vision Transformer to learn knowledge from discovered diverse domains. To facilitate cross-domain knowledge learning between different prompts, we introduce a domain prompt generator that enables knowledge sharing between domain prompts and a shared prompt. A domain mixup strategy is additionally employed for more flexible decision margins and mitigates the risk of incorrect domain assignments. Extensive experiments on three medical image classification tasks and one debiasing task demonstrate that our method can achieve comparable or even superior performance than conventional DG algorithms without relying on domain labels. The code is available at https://github.com/SiyuanYan1/PLDG.
\end{abstract}

\section{Introduction}
\label{sec1}


\begin{figure}[!t]
{\includegraphics[width=1\linewidth]{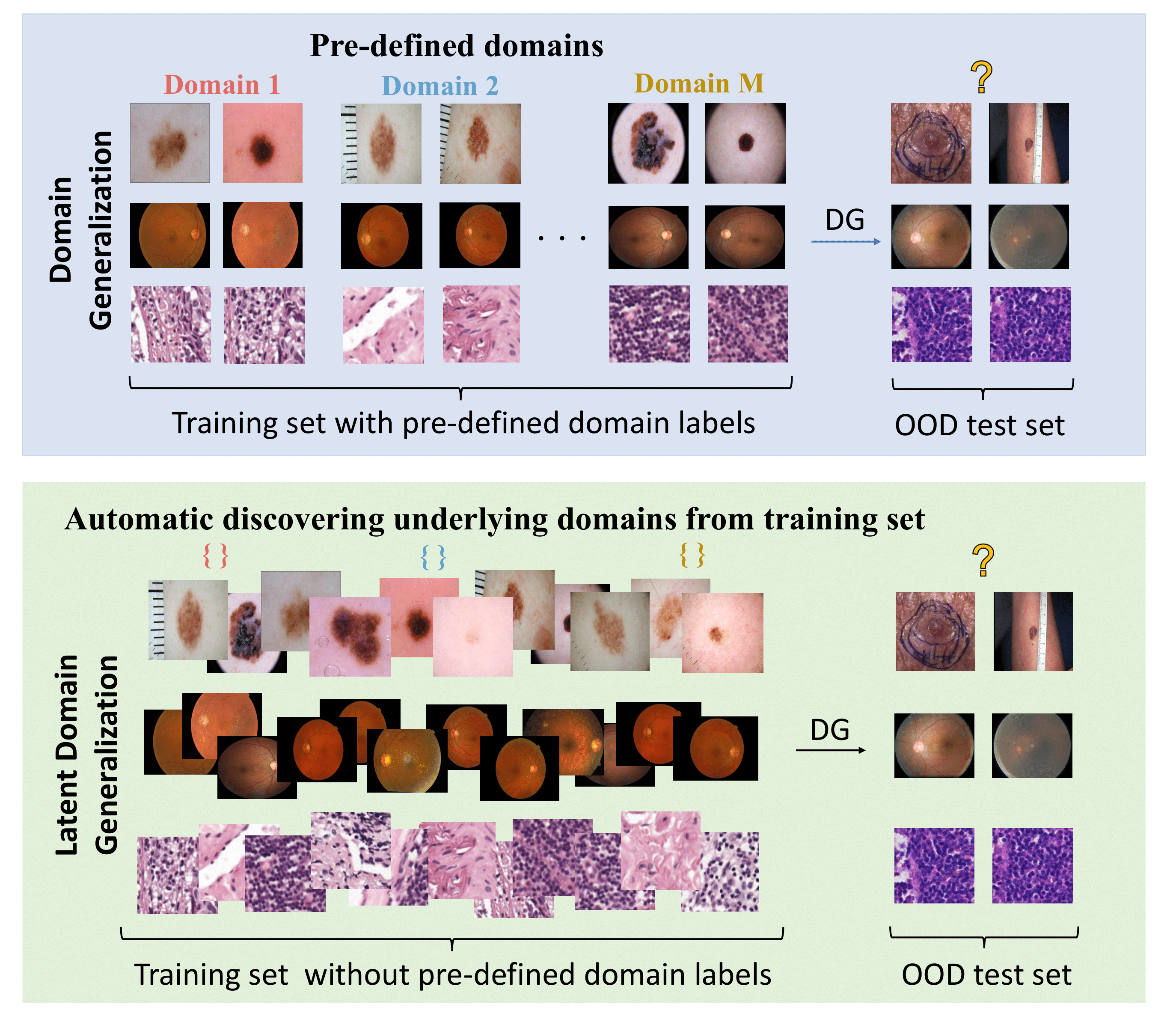}}
\caption{The comparison between conventional domain generalization (DG) and our latent domain generalization. Conventional DG aims to train the model to learn from multiple
domains to generalize well in unseen domains. Latent domain generalization aims to automatically discover essential domain information from a training set, enabling the training of a DG algorithm capable of generalizing to unseen domains.} 
\label{fig1}
\vspace{-3mm}
\end{figure}

Deep learning has made remarkable progress in various applications. However, recent studies \cite{shortcut2,shortcut3} have highlighted the vulnerability of deep learning models against distribution shifts caused by dataset bias, camera variations, differences in the imaging station, etc., which poses a significant risk in life-critical scenarios such as medical image analysis. For example, in skin cancer diagnosis using dermoscopic images, models may excessively rely on ``dermoscopic artifacts" such as rulers, gel bubbles, dark corners, and hairs \cite{artifacts,debias1,trustderm}, rather than learning the correct lesion patterns, leading to unreliable diagnoses. Similarly, models for Diabetic Retinopathy (DR) classification in ophthalmology can overfit specific camera styles, rendering them ineffective on novel images with different style features \cite{dr_gen,dr2}. This overreliance on specific cues rather than learning real patterns hinders the models' performance in real-world clinical environments where such cues are absent or inconsistent.

 A number of methods have been proposed to alleviate this issue from the perspective of domain generalization (DG). DG aims to train models on multiple different but related domains and expects them to perform well on unseen test domains. For instance, illustrated in the upper part of Fig. \ref{fig1}, the domains within medical datasets can be various factors, including dermoscopic artifacts such as hairs, rulers, and dark corners for skin cancer diagnosis, distinct camera devices for diabetic retinopathy (DR) classification, and diverse hospitals for histopathology images. However, DG sitll remains underexplored and relatively ineffective in medical image analysis due to several reasons from different perspectives. From a dataset-centric perspective: 1) domain labels in medical datasets are often unavailable as they are expensive and laborious to acquire; 2) the definition of visual domains in medical datasets is more ambiguous than natural images that can be clearly defined (e.g., photo, art painting, sketch, and cartoon in the PACS \cite{pacs} dataset). Specialists differing in clinical experience, diagnostic interests, etc., may have varying opinions on the optimal domain definition, making it challenging to define domains accurately; 3) the domain splitting in medical datasets can be heavily associated with the downstream tasks, making it difficult to transfer DG algorithms from one medical task to another due to inconsistent domain labels. From an algorithm-centric perspective, previous DG algorithms that learn domain-invariant features can also cause to ignore signals that are useful for unseen novel domains \cite{ermpp}. Although ensemble learning methods based on domain experts \cite{norm,domain_adap} can mitigate this limitation by learning domain-specific knowledge from different source domains independently, they overlook the rich cross-domain information that all domain experts can collectively contribute to the target domain prediction.


To address the domain challenges specific to medical image analysis, in this paper, we reconceptualize the DG problem in medical datasets as latent domain generalization (LDG), where generalized models are desired to learn from multiple underlying medical domains without relying on any pre-defined domain labels, as shown in the lower part of Fig.\ref{fig1}. Correspondingly, we proposed a novel, universal, prompt-driven LDG framework, called PLDG (\textbf{P}rompt-driven \textbf{L}atent \textbf{D}omain \textbf{G}eneralization), to alleviate the above-mentioned challenges from both perspectives.

The proposed PLDG framework consists of unsupervised domain discovery and domain prompt learning. The unsupervised domain discovery module aims to address the dataset-centric LDG challenges. We propose to discover and cluster the implicit dataset biases utilizing the Simplicity Bias property of learning-based algorithms \cite{sb1,sb2,sb3}. The clustering is performed based on the style features extracted from the shallow layer of the Vision Transformer (ViT). As the style features contain the implicit cues of common medical biases such as artifacts, skin tone, and image style, their clusters can serve as pseudo-domain labels. To address the algorithm-centric problem, we propose an ensemble-like domain prompt learning strategy, which leverages multiple lightweight domain prompts to enhance the learning of domain-specific knowledge from diverse source domains. Unlike existing ensemble-like methods that learn domain knowledge independently, we introduce a domain prompt generator to enable different domain prompts to collaborate and benefit mutually via low-rank weight updating so as to facilitate cross-domain knowledge learning. Furthermore, we employ a domain mixup strategy to mitigate the problem of noisy domain label assignments caused by unsupervised domain clustering.

In this paper, we make the following contributions:

1) We present a novel framework called Prompt-driven Latent Domain Generalization (PLDG) to address domain generalization in medical image classification without the need for explicit reliance on domain labels.

2) We propose a novel Simplicity Bias-guided pseudo domain label discovery method for arbitrary medical datasets.

3) We propose a prompt-based DG algorithm that takes advantage of a ViT-based domain prompt learning strategy and a novel domain prompt generator to promote cross-domain knowledge learning.

4) We benchmark our LDG framework and compare it with extensive existing DG algorithms using ViT-based backbones on three medical tasks and one debiasing task. The results demonstrate that our method achieves comparable or even superior performance without relying on any domain labels.

The preliminary version of our work, presented at MICCAI 2023\cite{epvt}, introduced the first prompt-based domain generalization method for skin lesion recognition. However, similar to most DG algorithms, the previous work limited itself to a narrow domain generalization task specific to skin lesion datasets and required annotated dermoscopic artifacts as domain labels. In this paper, we have extended the original method from a conventional DG framework to a novel LDG framework, which is universal for different medical classification datasets without relying on domain labels. The main advancements include: 1) we incorporate a novel Simplicity Bias-guided clustering step, which can discover pseudo domain labels directly from the datasets, eliminating the requirement for pre-defined domain labels; 2) we make the framework applicable to diverse medical classification datasets, offering flexibility and adaptability across a wide range of scenarios; 3) we validate the effectiveness of our method in far more datasets and tasks, including the original skin lesion datasets (Dermatology), four fundus datasets (Ophthalmology) and the Camelyon17-wild (Histopathology) benchmark dataset; 4) we provide a thorough analysis of important components and hyper-parameters to ensure the stability of our approach; 5) we benchmark both DG and LDG algorithms across three distinct medical classification tasks to facilitate future research.

\section{Related Work}

\subsection{Domain Generalization}
Domain generalization focuses on learning models that can generalize well to unseen target domains despite distribution shifts. Previous approaches have focused on learning domain-invariant features. For instance, DANN \cite{dann} aligns feature distributions from different source domains using an adversarial loss, while CORAL \cite{coral} matches the second-order statistics of different source domains. Other methods utilize model ensembles to explicitly learn domain knowledge through different model parameters. DAS and DNM \cite{domain_adap,norm} learn an ensemble of multiple classifiers or batch normalization statistics for different source domains. DoPrompt \cite{doprompt} embeds extra prompts into the network to capture domain-specific knowledge independently. Additional techniques include meta-learning \cite{meta, meta2}, feature disentanglement \cite{sagnet}, data augmentation \cite{mixup2}, and distributional robust learning \cite{dro} for achieving domain generalization.

Another direction for domain generalization is to leverage the power of deep learning architectures to learn stronger representations. Sarath et al. \cite{study} demonstrate different architectures exhibit varying performance on domain generalization datasets, with the vanilla Empirical Risk Minimization (ERM) outperforming many state-of-the-art algorithms when benchmarked using ResNet-50. Additionally, Dosovitskiy et al. \cite{vit16} show transformer-based architectures generally outperform ResNet-50 on domain generalization datasets as they are less biased towards texture. In this paper, we extensively benchmark domain generalization algorithms using the Vision Transformer (ViT) backbone on medical domain generalization datasets and take advantage of ViT's design to develop our own prompt-driven domain generalization algorithm.

\subsection{Domain Generalization in Medical Images}

Compared to domain generalization in natural images, domain generalization in medical images has received relatively less attention, mainly due to the challenges in obtaining domain labels for medical datasets. Bissoto et al. \cite{artifacts} annotate artifact-based domain labels in skin datasets using multiple artifact classifiers, while Mohammad et al. \cite{dr_gen} combine four common Diabetic Retinopathy (DR) datasets to construct the largest benchmark for domain generalization in DR classification where domain labels reflect as different datasets. However, both approaches have limitations compared to domain generalization datasets in natural images. For instance, the domain labels in skin datasets are noisy as they are only annotated by binary classifiers rather than human annotators. The definition of domains in the DR dataset is also suboptimal, as images from EyePACS \cite{eyepacs} and APTOS \cite{aptos} datasets are captured by multiple different cameras. The domain label, in this case, is dataset difference. Therefore, there is a need for latent domain generalization methods that can infer reasonable pseudo domain labels for domain generalization in any medical classification dataset. To acheive it, our work clusters style-based features as pseudo-labels. Although Toshihiko et al. \cite{dmld} explored inferring style features by clustering convolutional feature statistics from CNN during each training iteration, there is no existing work that explores how to obtain style features from ViT architectures in an efficient way.

\subsection{Prompt Learning}
Prompt learning was originally designed for natural language processing and involved prepending heuristics (manually designed) or learnable prompts (continuous vectors) into the input text, enabling large language models to handle various downstream tasks. Recently, prompt learning has also been applied to computer vision tasks. VPT \cite{vpt} inserts a series of learnable randomly initialized prompts into the pre-trained ViT and optimizes these prompts for diverse downstream tasks using corresponding task labels. Wang \etal \cite{clp} incorporates prompt tuning methods into continual learning tasks, which leverages multiple learnable prompts to handle corresponding tasks. Doprompt \cite{doprompt} designs a series of learnable prompts for different domains to capture domain-specific knowledge independently for domain generalization. In contrast to existing methods, our prompt learning strategy introduces a novel domain prompt generator that enables different domain prompts to collaborate and learn from each other, explicitly enforcing prompts to learn cross-domain knowledge for target domain generalization.

\section{Method}
\begin{figure*}[!t]
   \begin{center}
   {\includegraphics[width=0.8\linewidth]
   {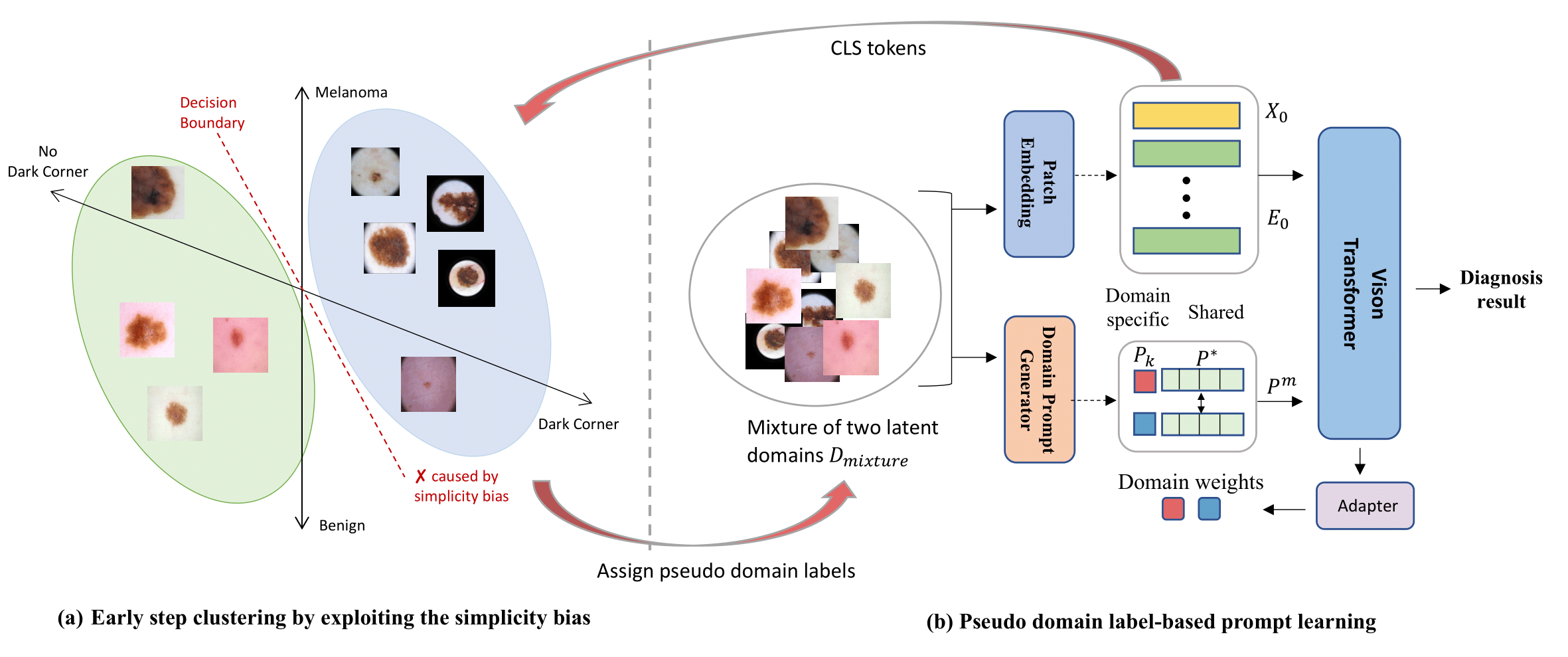}}
   \end{center}
  \vspace{-4mm}
\caption{\small 
Illustration of our prompt-driven latent domain generalization (PLDG) algorithm, (a) We perform one-time clustering on the CLS token from the shallow layer of the ViT model to discover the bias-related pseudo domain labels (see \ref{seca}). (b) Train a domain prompt-based ViT to learn domain-specific knowledge for unseen domain prediction (see \ref{secb}). A domain prompt generator is further employed to facilitate cross-domain knowledge learning (see \ref{secc}).
}
\label{overview}
   
\end{figure*}

 In conventional domain generalization (DG), the training dataset $D_{train}$ consists of $M$ source domains, denoted as $D_{train}=\lbrace D_{k}| k=1,...,M\rbrace$. Here, each source domain $D_{k}$ is represented by $n^{k}$ labeled instances $\lbrace(x_{j}^{k},y_{j}^{k})\rbrace_{j=1}^{n^{k}}$, which can also be represented by $D_{train}=\lbrace (x_{i},y_{i},d_{i}) \rbrace_{i=1}^{n}$ where $d_{i}$ denotes the domain labels and $n$ denotes the total training instances. The goal of DG is to learn a model $G: X\rightarrow Y$ from the $M$ source domains so that it can generalize well in unseen target domains $D_{test}$. In latent domain generalization (LDG), unlike DG, the domain labels are unknown. The training dataset thus becomes $D_{train}=\lbrace (x_{i},y_{i}) \rbrace ^{n}_{i=1}$.

The overall pipeline of our prompt-driven latent domain generalization method, PLDG, is shown in Fig.\ref{overview}. Our method aims to learn a domain generalization model using pseudo-domain labels. To obtain style-based pseudo domain labels that capture spurious correlations in the data, we cluster the class token in the shallow layer of a ViT model during the early epochs, as described in Section \ref{seca}. By utilizing the pseudo-domain labels, we propose using domain-based prompts to learn domain-specific knowledge in Section \ref{secb}. Furthermore, we encourage cross-domain knowledge learning and alleviate the noisy domain label assignments problems using a domain prompt generator and domain-based Mixup in Sections \ref{secc} and \ref{secd}.

\subsection{Simplicity Bias-guided Pseudo Domain Label Clustering}
\label{seca}

In learning-based algorithms, there is often a tendency to overfit to biases towards simplicity, such as focusing on the background rather than the target lesion in medical images or capturing dermoscopic artifacts instead of skin lesions \cite{sb1,sb2,sb3,sb4}, as known as Simplicity Bias. We leverage the Simplicity Bias property to identify bias attributes that are highly correlated with the target class. We then use these attributes as pseudo-domain labels. To identify the bias attributes, we follow the approach of previous works \cite{sb5, sb6} and cluster the model's easy-to-learn biased features within each class. To ensure that the clustering captures domain-related features rather than category-related features, we leverage style features, as they provide a more discriminative cue for common medical biases, such as distinct color styles caused by different cameras or hospitals, dermoscopic artifacts and skin tone. Although existing works \cite{sb5} utilize convolutional statistic features as style features to perform clustering, there is limited research on applying this approach to ViT architectures. Inspired by ViT-based style transfer algorithms, Tumanyan \etal \cite{stylevit} propose an appearance loss $\mathcal{L}_{app}$ to align the style features of ViT between the generated $I_{t}$ and appearance image $I_{a}$ via the CLS token:
\begin{equation}
\mathcal{L}_{app}=\lVert T^{L}_{cls}(I_{t})-T^{L}_{cls}(I_{a})\rVert_{2}
\end{equation}

where $T^{L}_{cls}$ is the CLS token extracted from layer $L$ of the ViT. In our task, we utilize the CLS token $T^{1}_{cls}$ from the shallow layers (e.g., block 1) of the ViT for one-time k-means clustering, as shown in Fig. \ref{overview}. This choice is motivated by the fact that the shallow layers provide global appearance and style information \cite{stylevit,vit16}, as also demonstrated in section \ref{cluster_exp}. Additionally, we find that performing one-time clustering in the early epochs is sufficient and yields stable results for discovering the pseudo-domain labels, as demonstrated in Section \ref{cluster_exp}. Once we obtain these pseudo-domain labels, we can apply domain generalization algorithms based on them to improve the model's performance on unseen data.

\subsection{Domain-specific Prompt Learning with Vision Transformer}
\label{secb}
\textbf{Domain-specific learning:} To enable the pre-trained vision transformer (ViT) to capture knowledge from different domains, we define a set of $M$ learnable domain prompts produced by a domain prompt generator (introduced in \ref{secc}), denoted as $P_{D}=\lbrace P^{m} \in \mathbb{R}^{d} \rbrace_{m=1}^{M}$, where $d$ is the same size as the feature embedding of the ViT, and each prompt $P^{m}$ corresponds to one specific domain. To incorporate these prompts into the model, we follow the conventional practice of visual prompt tuning \cite{vpt}, which prepends the prompts $P_{D}$ into the first layer of the transformer. Particularly, for each prompt $P^{m}$ in $P_{D}$, we extract the domain-specific features as:
\begin{equation}
F_{m}(x) = F([\ X_{1}, P^{m}, E_{1}\ ])
\end{equation}
where $F$ is the feature encoder of the ViT, $X_{1}$ denotes the class token, $E_{1}$ is the image patch embedding, $F_{m}$ is the feature extracted by ViT with the $m$-th prompt, and $1$ is the index of the first layer. Domain prompts $P_{D}$ are a set of learnable tokens, with each prompt $P^{m}$ being fed into the vision transformer along with the image and corresponding class tokens from a specific domain; the domain-specific prompt optimization is defined as:

\begin{equation}
\label{eq3}
    \mathcal{L}_{domain}= \mathcal{L}_{CE}(H({F}_{m}(x)),y)
\end{equation}

where $H$ is the classification head, $\mathcal{L}_{CE}$ is the cross entropy loss. Through optimizing, each prompt $P^{m}$ becomes a domain expert only responsible for the images from its own domain. By the self-attention mechanism of ViT, the model can effectively capture domain-specific knowledge from the domain prompt tokens.

\begin{figure}[!t]
\centering
\includegraphics[width=0.48\textwidth]{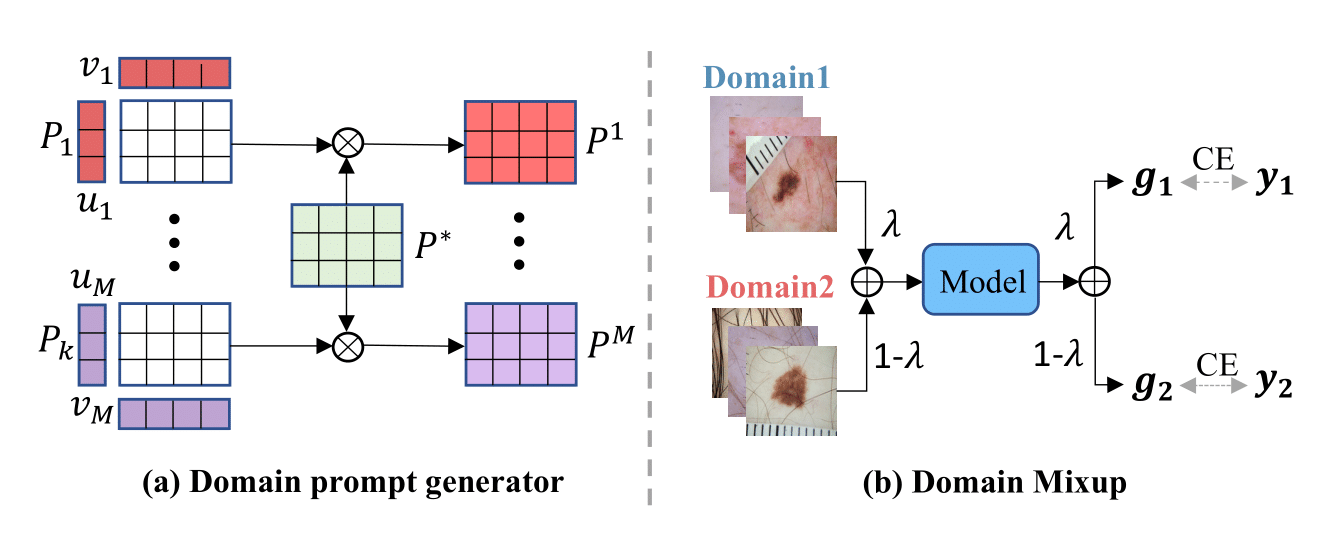}
\caption{Illustration of (a) domain prompt generator and (b) domain Mixup strategy.}
  \vspace{-2mm}
\label{f3}
\end{figure}

\textbf{Domain Mixup:} While optimizing $\mathcal{L}_{domain}$ based on the pseudo-domain labels inferred by clustering, a challenge arises due to the potential incorrect assignments of domains. To alleviate this issue, we propose employing the domain mixup strategy \cite{mixup2} on the domain loss term ($\mathcal{L}_{domain}$) to leverage inter-domain information. Instead of assigning a binary label ("0" or "1") to each image, a mixing operation is applied to every image in each batch. This mixing operation involves randomly selecting two images from different domains and combining them. As shown in Fig. \ref{f3}.b, the loss function $\mathcal{L}_{mixup}$ is then computed based on the predictions of the mixed images and their corresponding labels:
\begin{equation}
\begin{split}
\mathcal{L}_{mixup}&=\lambda \mathcal{L}_{CE}(H(F_{m}(x_{mix})),y_{i})\\&+(1-\lambda) \mathcal{L}_{CE}(H(F_{m}(x_{mix})),y_{j})
\label{mixup}
\end{split}
\end{equation}
where $x_{mix}=\lambda x_{i}^{k}+(1-\lambda)x_{j}^{q}$; $x_{i}^{k}$ and $x_{j}^{q}$ are samples from randomly two different domains $k$ and $q$, and $y_{i}^{k}$ and $y_{j}^{q}$ are the corresponding labels. 

\subsection{Cross-domain Knowledge Learning using Domain Prompt Generator}
\label{secc}
To facilitate effective knowledge sharing across different domains while preserving the domain-specific parameters of each domain prompt, we introduce a domain prompt generator, as illustrated in Fig. \ref{f3}.a. Our approach draws inspiration from model adaptation and multi-task learning techniques used in natural language processing \cite{mtl,compacter}. Aghajanyan \etal \cite{low_rank} have demonstrated that when adapting a model to a specific task, the weight updates exhibit a low intrinsic rank. Similarly, each domain prompt $P^{m}$ should possess a unique low intrinsic rank when learning knowledge from its respective domain. To achieve this, we decompose each $P^{m}$ into a Hadamard product between a randomly initialized shared prompt $P^{*}$ and a rank-one matrix $P_{k}$ obtained from two randomly initialized learnable vectors $u_{k}$ and $v_{k}$, given by:

\begin{equation}
P^{m} = P^{*} \odot P_{k} \quad \text{where} \quad P_{k} = u_{k} \cdot v_{k}^{T}
\end{equation}

Here, $P^{m}$ represents the domain-specific prompt, computed as the Hadamard product of $P^{*}$ and $P_{k}$. The shared prompt $P^{*} \in \mathbb{R}^{s \times d}$ is used to learn general knowledge, where $s$ and $d$ denote the dimensions of the prompt vector and feature embedding, respectively. On the other hand, $P_{k}$ is computed using domain-specific trainable vectors: $u_{k} \in \mathbb{R}^{s}$ and $v_{k} \in \mathbb{R}^{d}$. These vectors capture domain-specific information in a low-rank space. By decomposing the domain prompts into rank-one subspaces, the model can effectively learn domain information. The Hadamard product enables the model to leverage cross-domain knowledge for target domain prediction.

\subsection{Optimization and Inference}
\label{secd}

So far, we have introduced $\mathcal{L}_{mixup}$ in Eq. \ref{mixup} for optimizing our model. However, since our goal is to generalize the model to unseen environments, we need to take advantage of each domain prompt. Instead of assigning equal weights to each domain prompt, we employ an adapter \cite{doprompt} that learns the linear correlation between the domain prompts and the target image prediction.
To obtain the weighted prompt for inference in the target domain, we define it as a linear combination of the source domain prompts:
\begin{equation}
P_{weighted}=A(F(x))=\sum_{m=1}^{M}w_{m}\cdot P^{m} \textrm{,} \quad \textrm{s.t.} \quad \sum_{m=1}^{M}w_{m}=1
\end{equation}
where $A$ represents an adapter containing a two-layer MLP with a softmax layer, and $w_m$ denotes the learned weights. 

To train the adapter $A$, we simulate the inference process for each image in the source domain by treating it as an image from the pseudo-target domain. Specifically, we first extract features from the ViT: $\hat{F_{m}}(x)=F([X_{0},E_{0}])$. Then we predict the weighted prompt $P_{weighted}$ for the pseudo-target environment image x using the adapter $A$: $P_{weighted}=A(\hat{F_{m}}(x))$. Next, we extract features from ViT conditioning on the weighted prompt: $\hat{F_{m}}(x)=F([\hat{F_{m}}(x),P_{weighted},E_{0}])$. Finally, the classification head $H$ is applied to predict the label y: $y=H(\hat{F}_{m}(x))$. Additionally, the inference process is the same as the simulated inference process during adapter training, and our final prediction will be conditioned on the weighted prompt $P_{weighted}$. 

To ensure that the adapter learns the correct linear correlation between the domain prompts and the target image, we use the pseudo domain label from source domains to directly supervise the weights $w_{m}$. We also use the cross-entropy loss to maintain the model performance with the weighted prompt:
\begin{equation}
 \label{eq5}
 \footnotesize
\begin{split}
\mathcal{L}_{weighted}&=\mathcal{L}_{CE}(H(\hat{F}_{m}(x)),y) \\ &+ \lambda (\frac{1}{M}\sum_{m=1}^{M}\frac{1}{M}(\mathcal{L}_{CE}(w_{m}^{m},1) +\sum_{t\neq m}\mathcal{L}_{CE}(w_{t}^{m},0))
 \end{split}
\end{equation}

where $\hat{F}_{m}(x)$ is the obtained feature map conditioned on the weighted prompt $P_{weighted}$, and $H$ is the classification head. The total loss is then defined as $\mathcal{L}_{total}=\mathcal{L}_{mixup}+\mathcal{L}_{weighted}$.

\section{EXPERIMENTAL RESULTS}

\subsection{Experimental Setup}

\subsubsection{Dataset Description}

We evaluate the generalization ability of our method in four challenging classification settings that closely mimic real-world scenarios. Notably, our method stands out by not relying on domain labels. However, to facilitate a meaningful comparison with conventional domain generalization algorithms that require domain labels, we selected three domain generalization datasets for evaluation. It is important to note that during our evaluation, our method was assessed without utilizing any domain labels, while all other baseline algorithms were evaluated using domain labels.

1) \textbf{DG in Melanoma Classification:} We use the \textit{ISIC2019} dataset \cite{isic} for training and validation, which consists of melanoma and benign categories. The training set contains 12,360 images, and the validation set contains 2,060 images. In this setting, the domain labels are defined based on artifact annotations from \cite{artifacts}. The training set of \textit{ISIC2019} is divided into five groups: \textit{dark corner}, \textit{hair}, \textit{gel bubble}, \textit{ruler}, and \textit{clean}, with 2,351, 4,884, 1,640, 672, and 2,796 images, respectively. For testing, we use four out-of-distribution (OOD) datasets from \cite{artifacts}: \textit{Derm7pt-Dermoscopic} \cite{derm7pt}, \textit{Derm7pt-Clinical} \cite{derm7pt}, \textit{PH2} \cite{ph2}, and \textit{PAD-UFES-20} \cite{pad}. These datasets consist of 872, 839, 200, and 531 images, respectively. It is important to note that \textit{ISIC2019}, \textit{Derm7pt-Dermoscopic}, and \textit{PH2} are dermoscopic images, while \textit{Derm7pt-Clinical} and \textit{PAD} are clinical images. Model selection is performed using the training-domain validation set method \cite{dg}.

2) \textbf{DG in Diabetic Retinopathy (DR) Classification:} To evaluate OOD generalization in DR classification, followed by \cite{dr_gen}, we combine four commonly used DR datasets: EyePACs \cite{eyepacs}, Aptos \cite{aptos}, Messidor, and Messidor 2 \cite{messidor}. The combined datasets contain 35,126, 3,657, 1,200, and 1,744 images, respectively. The datasets consist of five categories, with grade 0 being the lowest form of DR and grade 4 being the most proliferative. Following \cite{dr_gen}, we train and validate our model on three datasets and test it on the remaining one. We report the testing results on all four datasets using this method.

3) \textbf{DG in Cancerous Tissue Detection:} The dataset CAMELYON17-WILDS \cite{wild} comprises histopathology images captured across different hospitals. Each image represents a 96x96 patch from a whole-slide image (WSI) of a lymph node section from a patient with potentially metastatic breast cancer. The category label indicates whether the patch contains a tumor, and the domain labels correspond to the five hospitals. The training set consists of 302,436 patches from 30 WSIs belonging to the first three hospitals. The validation set contains 34,904 patches from the fourth hospital, and the testing set contains 33,560 patches from the fifth hospital. Model selection is performed using the OOD validation method \cite{wild}.

4) \textbf{Debiasing in Skin Datasets:} We use the trap skin dataset \cite{artifacts} that contains seven artifacts. The dataset consists of six trap sets with increasing bias levels, ranging from 0 (randomly split training and testing sets from the ISIC2019 dataset) to 1 (the highest bias level where the correlation between artifacts and class label is in the opposite direction in the dataset splits). As the bias factor increases, the distribution difference caused between the training and testing sets also increases.
\begin{table}[!t]
\centering
\caption{The comparison results on four out-of-distribution melanoma classification datasets (ROC-AUC)}
\label{tab1}
\begin{threeparttable}
{\begin{tabular}{c|ccccc}

\hline
Method & DM7\_D & DM7\_C & PAD & PH2 & Average \\ \hline 
ERM  &$80.23$& $72.00$ & $75.74$ & $84.64$ & $78.15$ \\ 
DRO \cite{dro} & $82.55$ & $72.86$ & $80.02$ & $84.97$ & $80.10$ \\  
CORAL \cite{coral} & $80.12$ & $71.24$ & $\underline{88.17}$ & $86.98$ & $81.62$ \\ 
MMD \cite{inv2} & $81.40$ & $71.34$ & $84.95$ & $87.12$& $81.20$ \\ 
DANN\cite{dann} & $81.46$ & $72.07$ & $83.94$ & $85.94$ & $80.85$ \\  
IRM \cite{IRM} & $77.00$ & $70.21$ & $74.847$ & $78.84$ & $75.13$ \\ 

MLDG\cite{meta}& $82.94$ & $68.57$ & $78.59$ & $88.14$ & $79.56$ \\ 
CAD \cite{cad} & $82.72$ & $69.57$ & $81.36$ & $88.4$ & $81.51$ \\ 
DoPrompt \cite{doprompt} & $82.38$ & $71.61$& $83.81$& $\underline{91.33}$ & $82.06$ \\ 
SelfReg \cite{selfreg}& $81.43$ & $\underline{73.18}$ & $85.78$ & $89.28$ & $82.42$\\ 
MMLD\dag \cite{dmld} &$79.6$&$69$&$88.8$&$81.3$&$79.68$\\
\rowcolor{Gray}
EPVT \cite{epvt} & $\underline{83.25}$ &$\textbf{74.52}$ & $87.41$& $\textbf{92.53}$& $\textbf{84.43}$\\ 
\rowcolor{Gray}
PLDG\dag (Ours) & $\textbf{83.69}$ &$72.03$ & $\textbf{89.92}$& $89.09$& $\underline{83.68}$\\ 
\hline
\end{tabular}
\begin{tablenotes}
    \footnotesize
     \item [*]  \dag indicates the DG algorithm without domain labels.
    \item [*] Bold indicates the best result.
    \item [*] Underline indicates the second-best result.
    
\end{tablenotes}
}
\end{threeparttable}
\end{table}

\begin{table}[!t]
\centering
\caption{The comparison results on out-of-distribution Diabetic Retinopathy classification datasets (Acc)}
\label{tab2}

\begin{threeparttable}
{\begin{tabular}{p{1.72cm}|p{0.9cm}p{0.9cm}p{0.9cm}p{0.9cm}p{0.9cm}}

\hline
Method & EyePACS & APTOS & Messidor & Messidor2 & Average \\ \hline 
ERM  &$74.53$& $71.44$ & $57.75$ & $60.03$ & $65.94$ \\ 
CORAL \cite{coral} & $75.21$ & $71.65$ & $56.83$ & $62.27$ & $66.49$ \\ 
Fishr \cite{fishr} & $74.53$ & $\underline{71.68}$ & $57.83$ & $59.62$ & $65.92$\\
DANN\cite{dann} & $\underline{75.38}$ & $68.32$ & $54.92$ & $\underline{64.6}$ & $65.81$ \\  
SelfReg \cite{selfreg}& $\textbf{75.98}$ & $69.41$ & $\textbf{58.5}$ & $62.06$ & $66.49$\\
DoPrompt \cite{doprompt} & $73.04$ & $71.33$& $56.25$& $62.5$ & $65.78$ \\  
MMLD \dag \cite{dmld}&$71.32$&$66.5$&$55.12$&$61.39$&$63.58$\\
\rowcolor{Gray}
EPVT \cite{epvt} & $74.59$ &$71.57$ & $55.58$& $64.45$& $\underline{66.52}$\\ 
\rowcolor{Gray}
PLDG \dag (Ours) & $73.8$ &$\textbf{73.32}$ & $\underline{57.97}$& $\textbf{65.22}$& $\textbf{67.58}$\\ 
\hline
\end{tabular}

}
\end{threeparttable}
\end{table}
\subsubsection{Implementation Details} 
For a fair comparison, we use the ViT-Base/16 \cite{vit16} backbone, pre-trained on the ImageNet, as the base model for all experiments. The evaluation metrics include Accuracy for DR classification and the Camelyon17-WILDS dataset, and ROC-AUC for all other datasets. Hyperparameters play a crucial role in domain generalization algorithms, we conduct a grid search over the following hyperparameters for all models: learning rate (ranging from $3e^{-4}$ to $5e^{-8}$), weight decay (ranging from $1e^{-2}$ to $1e^{-5}$), and prompt length (ranging from 4 to 16, when available). We report the best performance achieved among all models. After the grid search, we employ the AdamW optimizer with specific hyperparameter settings for each task. For melanoma classification, we set the learning rate to $5e^{-6}$, weight decay to $1e^{-2}$, and the prompt length to 4. For cancerous tissue detection, we use a learning rate of $5e^{-6}$, weight decay of $1e^{-4}$, and a prompt length of 10. For DR classification, we use a learning rate of $5e^{-7}$, weight decay of $1e^{-5}$, and a prompt length of 4. All input images are resized to $224\times224$. Standard data augmentation techniques, such as random flip, crop, rotation, and color jitter, are applied. To prevent overfitting, we employ early stopping with patience of 22. All models are trained for a total of 60 epochs for out-of-distribution (OOD) evaluation and 100 epochs for trap set debiasing. The experiments are conducted on two NVIDIA RTX 3090 GPUs.
\subsubsection{Baseline Methods}
We compare our method with representative strong domain generalization baselines from the domainbed codebase \cite{dg}. These baselines cover various approaches in the domain generalization literature, including domain invariant representation learning \cite{IRM,dann,coral}, distributionally robust optimization \cite{dro}, feature disentanglement \cite{sagnet}, ensemble learning \cite{doprompt}, meta-learning \cite{meta}, gradient operation \cite{fishr}, prompt learning \cite{doprompt}, self-supervised learning \cite{selfreg}, latent domain adversarial learning \cite{dmld} and others \cite{cad}. To ensure a fair comparison, we benchmark all algorithms using the same ViT-Base/16 backbone, except for latent domain adversarial learning \cite{dmld}, which uses ResNet50 as it is specifically designed for convolutional neural networks. Additionally, we denote our method that utilizes domain labels (without the clustering step in section \ref{seca}) as EPVT, which corresponds to our conference version \cite{epvt}. Furthermore, we refer to our method without using domain labels as PLDG, which represents the latent domain generalization method proposed in this work.

\begin{table}[!t]
\centering
\caption{The comparison results on hospital five on Cancerous Tissue Detection datasets}
\label{tab3}

{\begin{tabular}{c|c}

\hline
Method & Accuracy\\ \hline 
ERM  &$73.1$ \\ 
CORAL \cite{coral} & $71.8$ \\ 
DANN\cite{dann} & $83.5$  \\  
IRM \cite{IRM} & $75$ \\ 
 
SelfReg \cite{selfreg}& $70.4$ \\ 
MMLD \dag \cite{dmld}&$70.2$\\
\rowcolor{Gray}
EPVT \cite{epvt} & $\textbf{86.4}$ \\ 
\rowcolor{Gray}
PLDG \dag (Ours) & $\underline{84.3}$ \\ 
\hline
\end{tabular}}
\end{table}

\subsection{Comparisons with existing domain generalization methods}
\subsubsection{Melanoma classification and Cancerous tissue detection}

\begin{figure*}[!t]
   \begin{center}
      \begin{tabular}{ c@{ } c@{ } }
  {\includegraphics[width=0.48\linewidth]{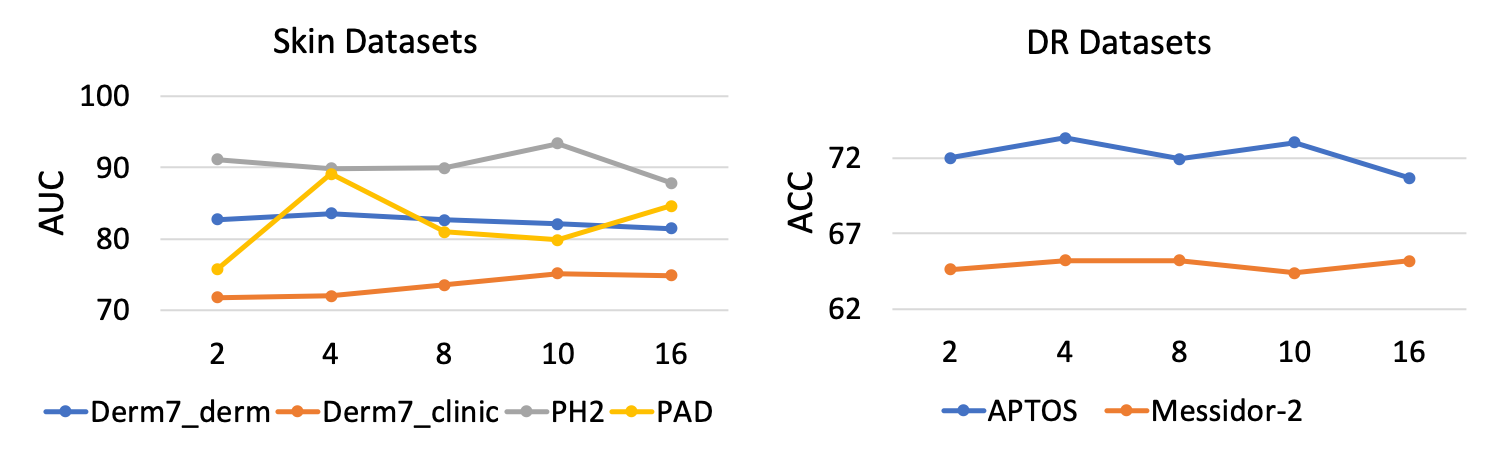}}&
  {\includegraphics[width=0.48\linewidth]{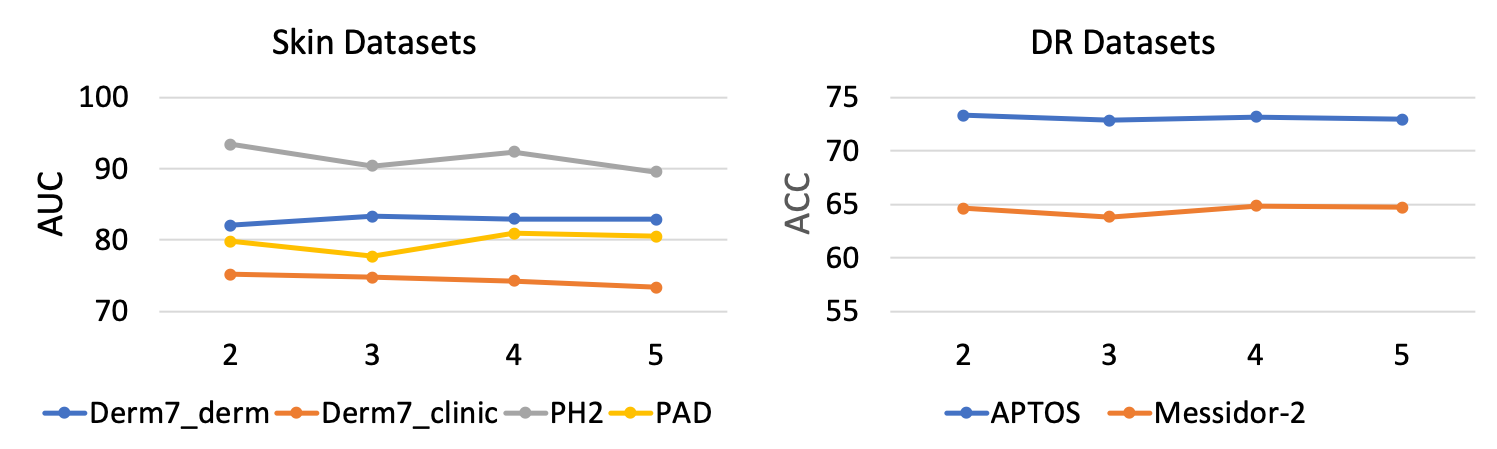}}\\
\footnotesize{(a) prompt length }& \footnotesize{(b) cluster number} 
    \end{tabular}
   \end{center}
      \vspace{-2mm}
\caption{\small Ablation analysis of (a) prompt length and (b) cluster number on six datasets of two tasks.}
   \label{fig4}
\end{figure*}

\begin{figure*}[!t]
   \begin{center}
      \begin{tabular}{ c@{ } c@{ } }
  {\includegraphics[width=0.48\linewidth]{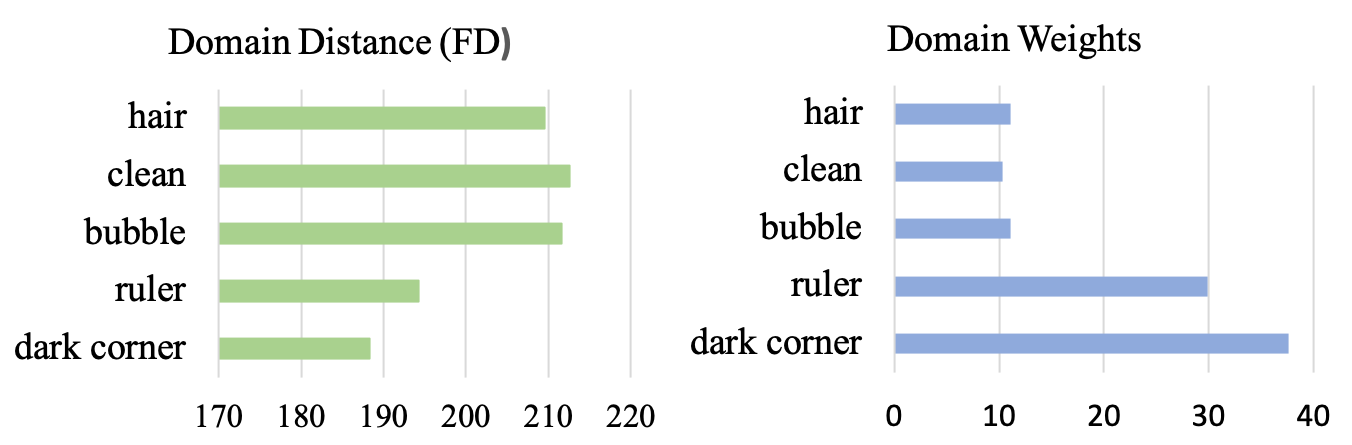}}&
  {\includegraphics[width=0.48\linewidth]{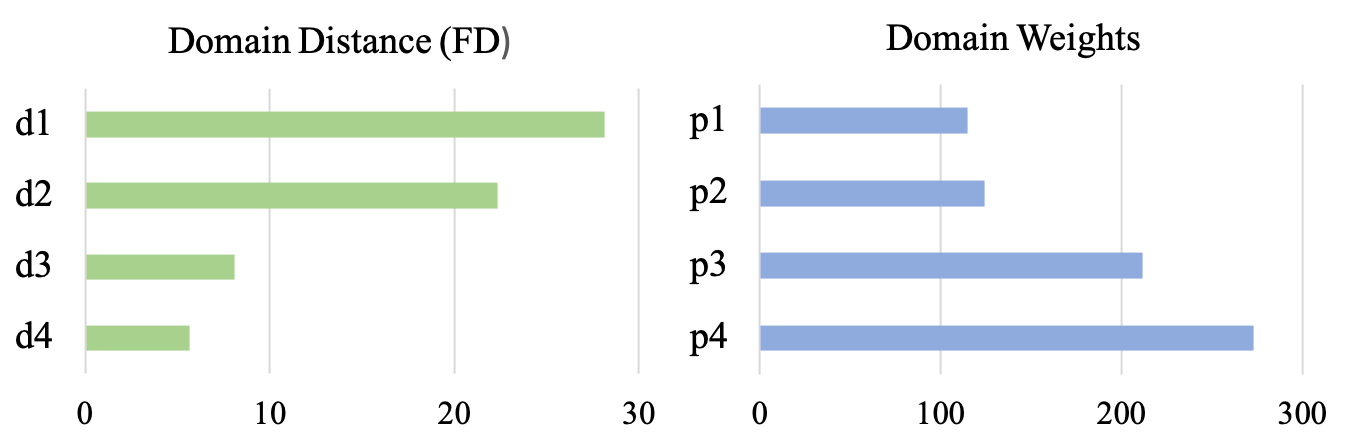}}\\
  \footnotesize{(a) w/ domain labels }& \footnotesize{(b) wo/ domain labels }
    \end{tabular}
   \end{center}
      \vspace{-2mm}
\caption{\small The relationship analysis of domain prompt weights and domain distance for our method with domain labels (a) and without domain labels (b).}
   \label{fig5}
\end{figure*}
Table \ref{tab1} and Table \ref{tab3} present a comparison of our PLDG algorithm with existing domain generalization methods, also including our algorithm with domain labels (EPVT) on Melanoma classification and Cancerous Tissue detection. The results clearly demonstrate the superiority of our approach. In melanoma classification, our PLDG algorithm achieves the best performance on two out of four OOD datasets and shows remarkable improvements over the ERM algorithm. Specifically, we achieve a 3.46\% improvement on the \textit{Derm7pt\_derm} dataset and a significant 14.18\% improvement on the \textit{PAD} dataset. Our algorithm also outperforms most state-of-the-art domain generalization algorithms on average performance and achieves competitive average performance with our method that utilizes domain labels. Similarly, our PLDG achieves the second-best performance on Cancerous Tissue Detection datasets, only performing worse than our method using domain labels. These results highlight the practicality of our latent domain generalization algorithm in the medical setting, as it shows comparable performance with the best domain generalization algorithm while eliminating the requirement for domain labels. This performance of our method also emphasizes the effectiveness of our prompt learning strategy in learning robust features for detecting melanoma and cancerous tissues, showcasing its potential in medical applications.
\begin{table}[!t]
\centering
\caption{Ablation study of PLDG on four OOD skin datasets}
\label{tabab1}
{\begin{tabular}
{c|ccccc}
\hline
Method& DM\_D & DM\_C & PAD & PH2 & Average \\ \hline 
baseline  &$80.23 $& $72.00$ & $75.74$ & $84.64$ & $78.15$ \\  
+P& $81.93$ & $73.56$ & $82.82$ & $87.89$ & $81.55$ \\ 
+P+A& $83.05$ & $72.45$ & $84.95$ & $86.17$ & $81.67$ \\  
+P+A+M& $82.55$ & $\textbf{73.73}$ & $86.80$ & $86.61$ & $82.42$ \\  
+P+A+M+G& $\textbf{83.69}$ &$72.03$ & $\textbf{89.92}$& $\textbf{89.09}$& $\textbf{83.68}$\\  \hline
\end{tabular}%
}
\end{table}

\subsubsection{Diabetic Retinopathy classification}
Table \ref{tab2} shows the comparison of our method with existing DG methods on DR classification. It can be seen that our method achieves the best average performance, surpassing even our algorithm with domain labels (EPVT). This result is surprising but reasonable considering the characteristics of the DR dataset benchmark \cite{dr_gen}. Unlike the previous two tasks, the DR dataset benchmark is created by simply combining four different datasets, where the domain labels are assumed to represent style differences primarily caused by variations in cameras. However, the images from the EyePACS and APTOS datasets are captured using multiple types of cameras, which makes the dataset-based domain separation sub-optimal. Further, it can be seen that all conventional domain generalization algorithms that rely on domain labels do not significantly improve the ERM baseline performance. In contrast, our PLDG algorithm shows a significant improvement in performance. This indicates that latent domain generalization is more effective when domain labels are noisy or unavailable, as it can effectively capture and utilize the underlying latent domain of the data.
\subsection{Ablation Study}
\begin{table}[!t]
\centering
\caption{Ablation study of PLDG on DR classification datasets}
\label{tabab2}
{\begin{tabular}
{c|ccc}
\hline
Method &APTOS& Messidor2&Average \\ \hline 
baseline&$71.44$ &$60.03$&$65.74$ \\  
+P& $72.01$&$61.07$&$66.54$ \\ 
+P+A &$71.64$&$61.75$&$66.61$ \\  
+P+A+M &$72.43$&$63.15$&$67.79$\\  
+P+A+M+G&$\textbf{73.32}$&$\textbf{64.62}$&$\textbf{68.97}$\\  \hline
\end{tabular}%
}
\end{table}
\begin{figure*}[!t]
   \begin{center}
      \begin{tabular}{ c@{ } c@{ } c@{ } c@{ } }
  {\includegraphics[width=0.25\linewidth]{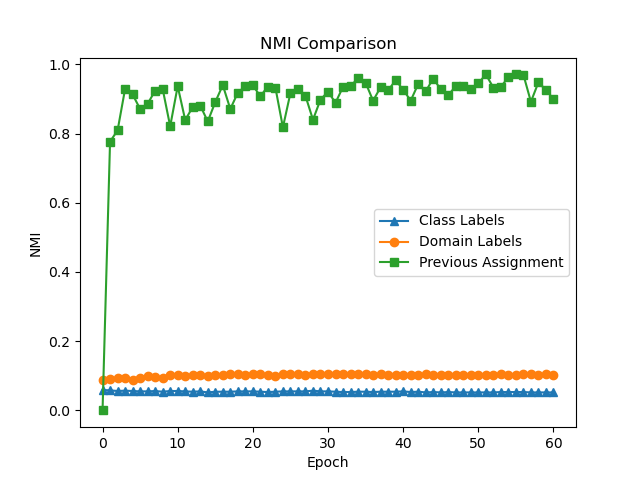}}&
  {\includegraphics[width=0.25\linewidth]{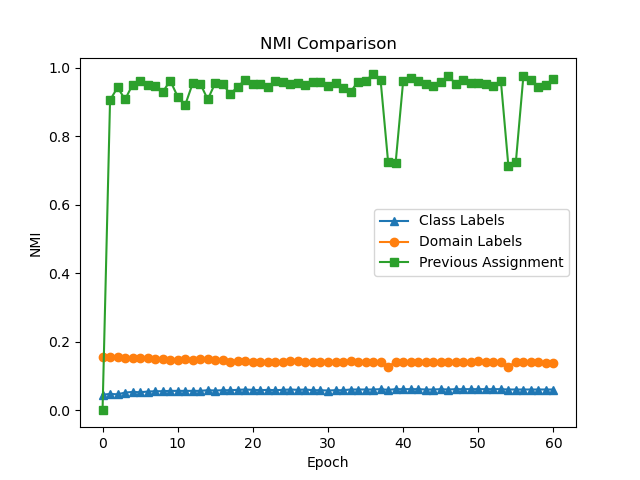}}&
    {\includegraphics[width=0.25\linewidth]{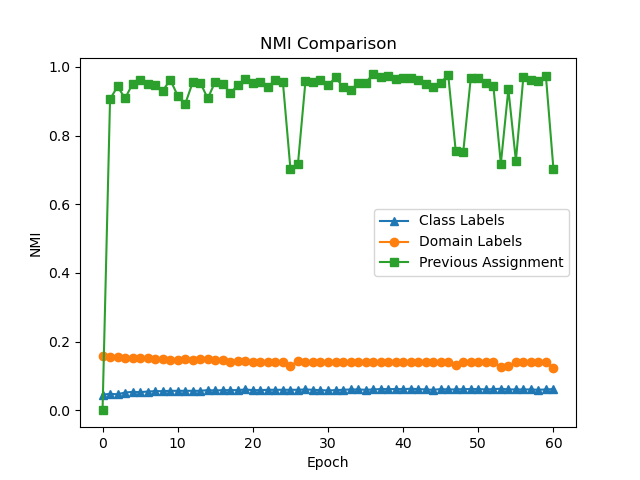}}&
  {\includegraphics[width=0.25\linewidth]{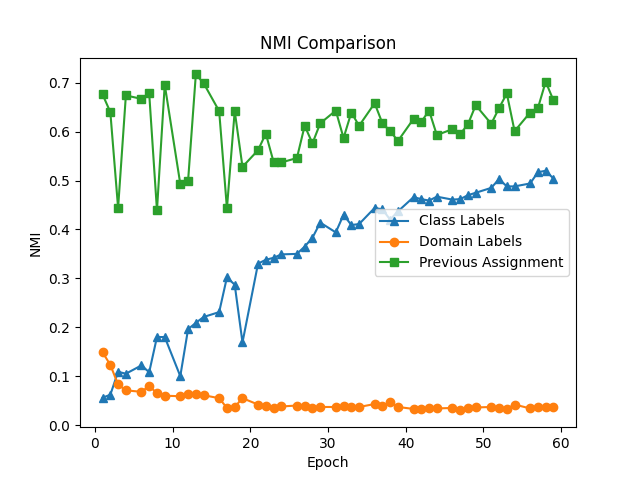}}\\
  \footnotesize{(a) L1 }& \footnotesize{(b) L5}&\footnotesize{(c) L8}&\footnotesize{(d) L12 }
    \end{tabular}
   \end{center}
      \vspace{-2mm}
\caption{\small Normalized mutual information between pseudo domain labels and category labels, pre-defined domain labels, and previous epoch assignments.}
   \label{fig6}
\end{figure*}

\subsubsection{Contribution of different components}
 We conduct ablation studies to analyze each component of our model, as shown in Table \ref{tabab1} and Table \ref{tabab2}. We set our baseline as the Empirical Risk Minimization (ERM) algorithm followed by conventional DG papers\cite{dg}, and we gradually add P (prompt \cite{vpt}), A (Adapter), M (Mixup), and G (domain prompt generator) into the model. For methods without an adapter, we use an equal-weighting mechanism. The performance of each component is evaluated on melanoma datasets and DR classification datasets. Firstly, we added multiple randomly initialized learnable prompts into the baseline and optimized them using the $\mathcal{L}_{\text{domain}}$ loss function in Equation \ref{eq3}, denoted as ``+P". Compared to the baseline, using the domain prompt learning strategy improved the average ROC-AUC by 3.39\% and the average accuracy by 0.8\% for the melanoma and DR datasets, respectively. This result demonstrates the effectiveness of the domain prompt learning component. Next, we incorporated the adapter and domain-based Mixup into the model, denoted as ``+P+A+M". Compared to "+P", the model achieved an average improvement of 0.87\% and 1.16\% on the melanoma and DR datasets, respectively. This finding highlights the importance of addressing wrong label assignments and utilizing domain weighting to improve generalization. Finally, we incorporated the domain prompt generator into the model, resulting in our PLDG approach, denoted as ``+P+A+M+G". It can be observed that combining the domain prompt generator improved the average ROC-AUC by 1.26\% and the average accuracy by 1.18\% on the two tasks. This emphasizes the importance of facilitating cross-domain learning in the context of medical domain generalization.
 
\subsubsection{Analysis on hyper-parameters}
In our method, the prompt length and cluster number are two important hyperparameters. We investigate the impact of different prompt lengths and cluster numbers on the performance of our method, and the results are shown in Fig. \ref{fig4}. For the prompt length, we find that setting it to 4 leads to the best average performance on both the skin and DR datasets. Moreover, when the prompt length is set to 10, our method achieves the best performance on specific datasets such as \textit{Derm7pt\_clinic} and \textit{PH2}. For the cluster number, we observe that setting it to 4 results in the best average performance for both datasets. Interestingly, we find that our method is not highly sensitive to the cluster number, as it consistently outperforms most domain generalization baselines when using different values of the cluster number ranging from 2 to 5.

\begin{figure}[t]
   \begin{centering}

  {\includegraphics[width=0.7\linewidth]{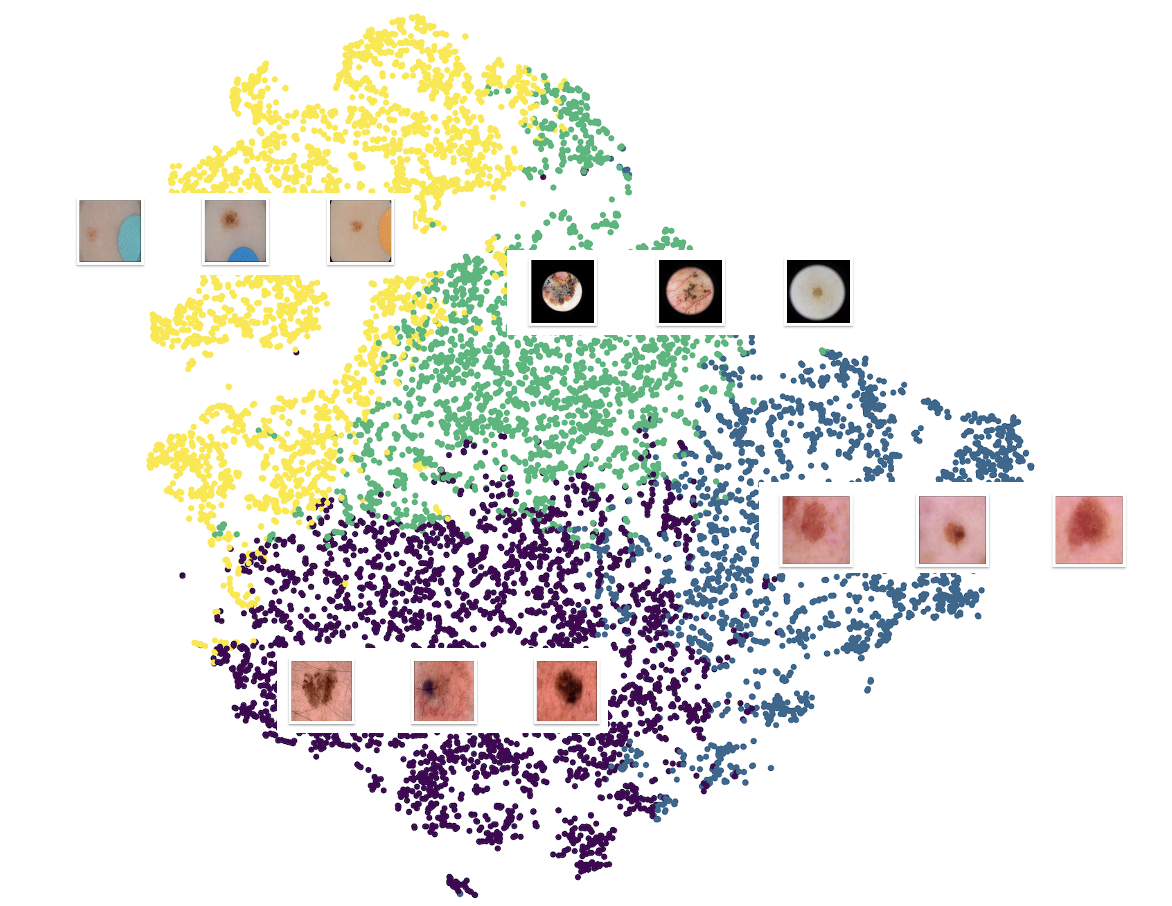}}
   \\
   \end{centering}
      \vspace{-2mm}
\caption{\small T-SNE visualization of pseudo domain labels.}
   \label{fig7}
\end{figure}

\subsection{Domain prompt weights analysis}
To evaluate whether our method has successfully learned the correct domain prompts for target domain prediction, we conduct an analysis and plot the results in Fig. \ref{fig5}. This analysis is performed for both our method with domain labels and our method without domain labels. Firstly, we extract the features of each domain (or pseudo domain using clustering) from our training set \textit{ISIC2019}, and we also extract the features from a target dataset, \textit{Derm7pt-Clinc}. Using these extracted features, we calculate the Frechet distance \cite{frechet} between each domain and the target dataset, which represents the domain distance between them. Next, we record the learned weights of each domain prompt. When domain labels are available, we observe that our model assigns the highest weight to the "dark corner" group, as the domain distance between the "dark corner" group and the \textit{Derm7pt-Clin} dataset is the closest, as shown in the right panel of Fig. \ref{fig5}(a). This indicates that the "dark corner" group shares the most similar domain information with the target dataset, and thus, it is given the highest weight. On the other hand, the "clean" group is assigned the smallest weight, as the domain distance between the "clean" group and the target dataset is the largest. This suggests that the domains of the "clean" group are significantly different from the target domain and contain less useful information for target domain prediction. A similar relationship can be observed when our model uses pseudo domain labels, as shown in Fig.\ref{fig5}.b. In both cases, there is a negative correlation between the domain distance and the corresponding prompt's weights. This implies that our model can precisely learn the relevant knowledge from different domains and assign higher weights to the domains that are more similar to the target domain.
\begin{figure}[t]
   \begin{centering}
  {\includegraphics[width=0.8\linewidth]{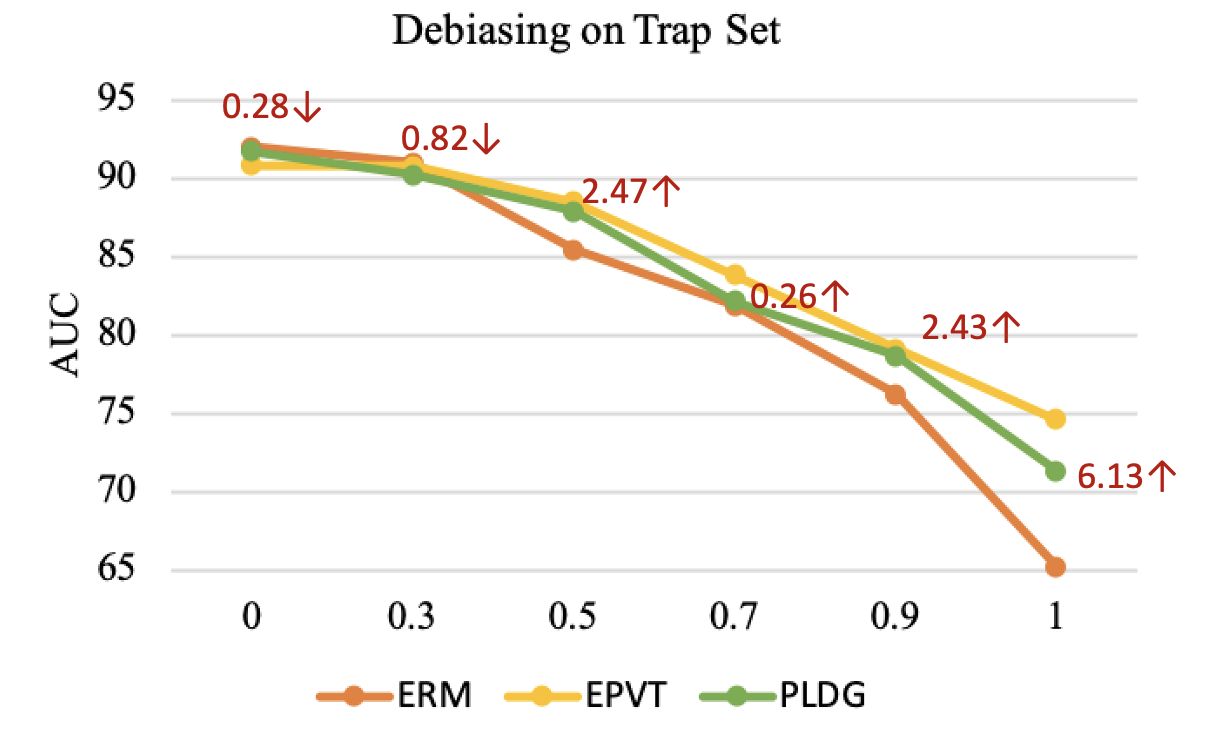}}
   \\
   \end{centering}
      \vspace{-2mm}
\caption{\small Debiasing evaluation on trap set consisting of six skin datasets with different bias levels.}
   \label{fig8}
\end{figure}
\subsection{Clustering analysis}
\label{cluster_exp}
Our method acquires pseudo domain labels via one-time clustering. One concern is whether the clustering is performed based on category labels rather than visual domains. To investigate this, we evaluated the correlation between the category labels and the pseudo domain labels. The results, shown in the blue lines in Fig. \ref{fig6}, indicate the normalized mutual information (NMI) between them at different layers (1, 5, 8, and 12) of the ViT. A high NMI value suggests a strong correlation. We observed that the correlation between pseudo domain labels and category labels is high in the last layer (L12) of the ViT but low in the shallow layer (L1). This indicates that the clustering was not performed based on category labels when selecting the CLS token from the shallow layer (L1). Furthermore, we noticed that the NMI between pseudo domain labels and their previous assignments became stable after a few epochs in layer 1. Based on these observations, we chose to cluster the CLS token from the later 1 in epoch 5, where the clustering is stable and not based on category labels.

However, Fig. \ref{fig6} also shows that the clustering is not performed totally based on the pre-defined domain labels in the original dataset, as evidenced by the relatively low NMI between the pseudo domain labels and the original domain labels. Although previous work by Deecke et al. \cite{sla} showed that the use of pre-defined domain labels in original datasets is not necessary, given that the domain labels in many well-known domain generalization datasets are sub-optimal, we are still curious about how our method clusters the samples. To gain further insights, we visualize the CLS token features in layer 1 for the skin dataset using T-SNE in Fig. \ref{fig7}. It reveals that the pseudo domain labels are still based on style features, with clusters representing "ink marking," "dark corner," "dark skin," and "light skin." This aligns with observations from the dermatology literature \cite{artifacts,color}. 


\subsection{Trap set debiasing}
In Fig. \ref{fig8}, we compare the performance of the ERM baseline, our method using domain labels (EPVT), and our method without domain labels (PLDG) on six biased trap datasets. Each point on the graph represents an algorithm trained and tested on a specific bias degree split of the trap set. The graph illustrates that the ERM baseline outperforms our PLDG when the bias degree is low (0 and 0.3). However, this can be attributed to the fact that ERM heavily relies on spurious correlations between artifacts and class labels, leading to overfitting the training set. As the bias degree increases, the correlation between artifacts and class labels decreases, and relying solely on artifacts for prediction becomes unreliable. This causes the performance of ERM to drop dramatically on the test set with a significant distribution difference. In contrast, our PLDG shows greater robustness to different bias levels. Notably, our PLDG outperforms the ERM baseline by 6.13\% on the bias 1 dataset. Although EPVT exhibits greater robustness than our PLDG, it is important to note that PLDG is more general for debiasing as it does not require domain labels for the dataset, making it applicable in real-world scenarios where domain labels are unavailable or noisy.

\section{CONCLUSION}

In this paper, we introduce a latent domain generalization method for medical image classification, and we try to answer some important questions: (1) whether domain labels are always necessary for medical domain generalization, and (2) whether the latent domain generalization method can outperform conventional domain generalization method that relies on domain labels.

To answer these questions, we propose a prompt-driven latent domain generalization framework that leverages pseudo domain labels obtained through clustering. Our extensive experimental results on different datasets have provided valuable insights. Firstly, we have shown that domain labels are not always necessary for achieving competitive performance in medical domain generalization tasks. Our method, without the use of domain labels, has achieved comparable performance to our method that employs domain labels and even outperformed most conventional SOTA domain generalization algorithms. This indicates that our method can effectively capture and leverage the underlying domain knowledge without explicitly relying on domain labels. Furthermore, our experiments have demonstrated that latent domain generalization methods can exhibit superior generalization abilities compared to conventional domain generalization methods, especially in scenarios where domain labels are either not available or not reliable. This highlights the practicality and versatility of our method in various medical settings, where obtaining precise domain labels can be challenging.

\bibliographystyle{IEEEtran}
\bibliography{IEEEabrv, main.bib}
\end{document}